\documentclass[apc,prc,twocolumn,superscriptaddress,showpacs,twoside,floatfix]{revtex4-1}
\usepackage{dcolumn}
\usepackage{bm}

\begin{document}

\title{Comment on ``Reviewing the evidence for two-proton emission from the high-spin
isomer in $^{94}$Ag''}

\author{I.~Mukha}
 \affiliation{CSIC -- IFIC Universidad de Valencia, E-46071 Valencia, Spain}
\author{E.~Roeckl}
  \affiliation{GSI Helmholtzzentrum  f\"{u}r Schwerionenforschung,
      D-64291 Darmstadt, Germany}
\author{H.~Grawe}
  \affiliation{GSI Helmholtzzentrum  f\"{u}r Schwerionenforschung,
      D-64291 Darmstadt, Germany}
\author{S.L.~Tabor}
  \affiliation{Florida State University, FL-32306 Tallahassee, USA}

\begin{abstract}
A recent publication [D.G.~Jenkins, Phys.~ Rev.~ C \textbf{80}, 054303 (2009)] claims to discredit the experimental observation
of two-proton decay of the (21$^+$) high-spin isomer in $^{94}$Ag [I.~Mukha \emph{et al.,} Nature (London)  \textbf{439}, 298 (2006)].
Its conclusion, which would require a reestablishment of the two-proton emission, is made on the basis of unwarranted assumptions
by Jenkins concerning the data analysis of the original work. We provide proof that
these assumptions do not correspond to reality, and that therefore the conclusion of the paper is misleading.

\end{abstract}


 \maketitle


The recent publication of Jenkins \cite{jenkins09} presents a very negative view of
the experiment reporting  the two-proton (2p) radioactivity from the (21$^+$) high-spin isomer $^{94}$Ag \cite{mukh_ag2p}.
The author challenges the unambiguous signature for 2p emission given in  \cite{mukh_ag2p}, namely the observation of 5 known
$\gamma$ rays from the lowest states in the 2p-decay daughter $^{92}$Rh. These $\gamma$ rays were registered in 4-fold coincidence
demanded between two silicon (Si) charged-particle and two germanium (Ge) $\gamma$-ray detectors (Si$_1$+Si$_2$+$\gamma_1$+$\gamma_2$). About 50000 decays of the (21$^+$) isomer in $^{94}$Ag were investigated in this way.
 In particular, Jenkins 
 claims that spurious peaks from Compton-scattered $\gamma$ rays associated with the dominant background from $^{94}$Ag $\beta$
decays could have been misidentified as $^{92}$Rh $\gamma$ rays. Such a claim is based on the author's unwarranted assumption that
the processes of $\gamma$-ray Compton scattering between adjacent Ge crystals were not reduced in the analyzed $\gamma$--$\gamma$
coincidences. However, the author has apparently overlooked that such a suppression procedure was applied as a standard routine
during the data analysis, as was mentioned in one of the preceding publications on  $^{94}$Ag $\beta$ decay, see page 28 in
Ref.~\cite{plett_ag_beta_g}.
This is surprising as Jenkins 
refers to previous publications on the same experiment considering different decay branches of isomers in $^{94}$Ag:
(i) $\beta$-delayed $\gamma$-ray emission \cite{plett_ag_beta_g}, (ii) $\beta$-delayed proton emission \cite{mukh_ag_beta_p},
and (iii) single-proton radioactivity \cite{mukh_ag1p}. As the author correctly points out, this series of papers reports on
successively weaker decay branches from the (21$^+$) isomer in $^{94}$Ag, obtained by analyzing and re-analyzing  the same data set.
The most complete description of the analysis used, the data obtained, and the calibrations are given in the two earlier regular
papers \cite{plett_ag_beta_g,mukh_ag_beta_p}. In particular, one of the basic routines used for reducing Compton scattering effects
in  $\gamma$--$\gamma$ coincidence events excluded double hits in adjacent Ge crystals while they were accepted in the other crystals \cite{plett_ag_beta_g}  (in addition to Ref.~\cite{mukh_ag2p}, we must bring that the coincidence events with the sum energy of two $\gamma$ rays amounting to 511$\pm$1.5 keV  were excluded for all crystals due to an association with positron annihilations).
This procedure has indeed reduced the Compton-scattered events, as one may conclude from the cross-check $\gamma$-ray spectrum shown
in Fig.~1(c) in Ref.~\cite{mukh_ag2p}. This spectrum was obtained by applying the same conditions as those used for projecting the
$\gamma$-ray spectrum with $^{92}$Rh evidence displayed in Fig.~1(d) of Ref.~\cite{mukh_ag2p}, except that the Si$_1$+Si$_2$ sum-energy gates were chosen
differently, i.e.\  covering the ranges of 1.2--1.6 and 1.8--1.95 MeV, respectively. If the effect of Compton scattering mocks-up
$^{92}$Rh $\gamma$-rays
as claimed by Jenkins then it should produce the same $\gamma$-ray peaks in both spectra mentioned above, which is clearly not
true. We are surprised that this straightforward cross-check published in \cite{mukh_ag2p} has escaped the attention of Jenkins.
 Another result of \cite{mukh_ag2p} questioned by Jenkins is the discrete 1.9-MeV Si$_1$+Si$_2$ sum-energy peak observed in
coincidence with two  $\gamma$ rays from excited states of the 2p-decay daughters $^{92}$Rh. The alternative interpretation by
Jenkins is that the reported 1.9 MeV peak was not produced by 2p but by electron-positron pairs generated by $\gamma$ decay of
the 2.86 MeV state in $^{94}$Rh which was assumed to be present in a sufficient amount due to $\beta$ decay of  $^{94}$Pd.
Besides ignoring the fact that $^{94}$Pd was greatly suppressed by using a cooling trap in the ISOL ion source and by selecting a
short collection-transport cycle of the tape station, this
interpretation contains a fancy qualitative assumption how two $\gamma$ rays de-exciting $^{92}$Rh states can be mocked-up by 4
511 keV $\gamma$-rays following annihilations of two positrons from $^{94}$Pd decay.
In particular, it assumes  that $\gamma$ rays from $^{92}$Rh can be simulated by Compton-scattered 511-keV photons.
Such a claim is not consistent with the above-mentioned fact that the data analysis used in \cite{mukh_ag2p} has excluded
$\gamma$--$\gamma$ hits with sum energy of 511~keV. In addition,  Jenkins has ignored the cross-check Si$_1$+Si$_2$ spectrum shown
in Fig.~1(b) of Ref.~\cite{mukh_ag2p} which has been projected by shifting coincident $\gamma$ gates by $\pm$3 keV from the nominal
$^{92}$Rh values. According to the interpretation of Jenkins, this spectrum should reveal the 1.9~MeV peak as well, in contrast to
the real data.
In this Comment, we put aside questions and problems of interpretation of the observed 2p decay. We believe that improved
measurements of the $^{94m}$Ag decay rather than wild guesses about the existing data will help in understanding of a physics
behind this phenomenon.

{\it In conclusion}, the Ref.~\cite{jenkins09} attempts to discredit the observed 2p decay of the (21$^+$) high-spin isomer in
$^{94}$Ag by using  wrong unwarranted assumptions about the data analysis applied in Ref.~\cite{mukh_ag2p}, and its unfounded
speculations contradict the two cross-check spectra given along with the data in \cite{mukh_ag2p}.

\end{document}